\documentclass[prb,twocolumn,superscriptaddress,a4paper,floatfix]{revtex4}
\usepackage{graphicx}
\usepackage{amssymb}
\usepackage{amsmath}
\usepackage{color}
\usepackage{paralist}
\usepackage{bm}
\usepackage{verbatim}

\def\hh{{\hat{h}}}

\def\bd{{\bf d}}

\def\bk{{\bf k}}
\def\bN{{\bf N}}
\def\bq{{\bf q}}

\def\bp{{\bf p}}
\def\br{{\bf r}}
\def\bR{{\bf R}}
\def\bS{{\bf S}}
\def\bv{{\bf v}}

\def\bnabla{{\boldsymbol \nabla}}

\def\hS{\hat S}
\def\hPsi{\hat \Psi}

\newcommand{\ket}[1]{| #1 \rangle}
\newcommand{\bra}[1]{\langle #1 |}

\newcommand{\Notes}[1]{{\color{blue}#1}}

\begin{document}

\title{Spin-orbital dynamics in a system of polar molecules}



\author{Sergey V. Syzranov}
\thanks{S.V.S. and M.L.W. contributed equally to this work.}
\affiliation{Department of Physics, University of Colorado, Boulder, CO 80309, USA}

\author{Michael L. Wall }
\thanks{S.V.S. and M.L.W. contributed equally to this work.}
\affiliation{JILA, NIST, and Department of Physics, University of Colorado, Boulder, CO 80309}

\author{Victor  Gurarie}
\affiliation{Department of Physics, University of Colorado, Boulder, CO 80309, USA}

\author{Ana Maria  Rey}
\thanks{Corresponding Author: {arey@jilau1.colorado.edu}}
\affiliation{JILA, NIST, and Department of Physics, University of Colorado, Boulder, CO 80309}

\date{\today}




\begin{abstract}
{\bf Spin-orbit coupling (SOC) in solids normally originates from the electron motion in the electric field of
the crystal. It is key to understanding a variety of
spin-transport and topological phenomena, such as
Majorana fermions and recently discovered topological insulators.
Implementing and controlling spin-orbit coupling is thus highly desirable and could open  untapped opportunities for the exploration of unique quantum physics.
Here, we show that  dipole-dipole interactions can produce an effective SOC in two-dimensional ultracold
polar molecule gases.  This SOC generates chiral excitations with a non-trivial Berry phase $2\pi$.
These excitations, which we call \emph{chirons}, resemble low-energy quasiparticles
in bilayer graphene and  emerge regardless of the quantum statistics and for arbitrary ratios of kinetic to interaction energies. Chirons manifest themselves
  in the dynamics of the spin density profile, spin currents, and spin coherences,
  even for molecules pinned in a deep optical lattice and should be observable in current experiments.

}
\end{abstract}

\date{\today}
\maketitle

Polar molecules~\cite{Carr_Review,Quemener_Julienne_12,barnett:quantum_2006,gorshkov:tunable_2011,gorshkov:quantum_2011,Wall_PRA_2010}  present a flexible platform for the exploration  of  quantum magnetism in many-body systems due to  their strong and long-range dipole-dipole interactions and their rich internal structure of rotational levels.
A few isolated rotational levels of a molecule represent
an effective spin degree of freedom.  Net spin-spin couplings can be directly generated by dipolar interactions even in frozen molecule arrays.
Recent experiments~\cite{Yan_Moses,Hazzard_Gadway_14} with  molecules pinned in a deep optical lattice
 have demonstrated dipolar spin exchange coupling. The anisotropic dipole-dipole interaction can also couple the spin degrees of freedom to the orbital motion of the molecules. Signatures  of this type of coupling have been recently reported experimentally in bosonic magnetic atoms~\cite{Pfau06,dePaz2013}, and have been noted for their potential  to generate topological phases\cite{NigelCooperReview,Manmana2013,Yao2013,PhysRevLett.110.145303}.
All previously predicted phenomena were limited to zero-dimensional systems\cite{Pasquiou2011} or
weakly interacting bosonic systems at zero temperature or
were tailored to particular experimental setups\cite{Sun2007,Santos2006,Gawryluk2007,Kawaguchi2006,Li_Wu},
requiring, e.g., complicated dressing techniques\cite{Manmana2013,Yao2013}.

Here, we demonstrate that an effective spin-orbit coupling (SOC) is inherent in the excitations of any two dimensional (2D) system
of polar molecules with a pair of degenerate $N=1$ rotational levels.
These excitations, which we call {\it chirons}, are characterised by a non-trivial Berry phase $2\pi$.
The same Berry phase is responsible for, e.g.,
 an unconventional quantum Hall effect in bilayer graphene\cite{McCannFalko:bilayer,Novoselov}.
Remarkably, in our system SOC emerges due to interactions rather than
being a single-particle effect, which adds significant richness to the physics and  removes the fundamental limitations imposed by spontaneous emission  present when single-particle SOC is
artificially generated by light \cite{Galitski2013,NigelCooperReview,Dalibard,PhysRevLett.111.185302,PhysRevLett.111.185301}.
We present ways to detect chirality and Berry phase, for instance, by
exciting rotational degrees of freedom in a finite spatial region.
Generically, this leads to two fronts
of spin and density currents,
corresponding to the two branches of the chiron spectrum.
The Berry phase $2\pi$ manifests itself in the {\it d-}wave symmetry $\propto\cos(2\phi)$
of the spin projection onto the plane, where $\phi$ is defined in Fig.~\ref{fig:Schematic}(a).
Additionally, the SOC leads to  population transfer between the excited rotational levels together with the
formation of a vortex structure in the spatial density profile.
We discuss the experimental conditions necessary for the observation of the described phenomena and provide
numerical examples germane to current polar molecule experiments in which molecules are pinned
in a deep optical lattice with sparse filling.

\section*{Many-body Hamiltonian for polar molecules in two dimensions}

\begin{figure}[h]
	\centering
	\includegraphics[width=1.0\columnwidth]{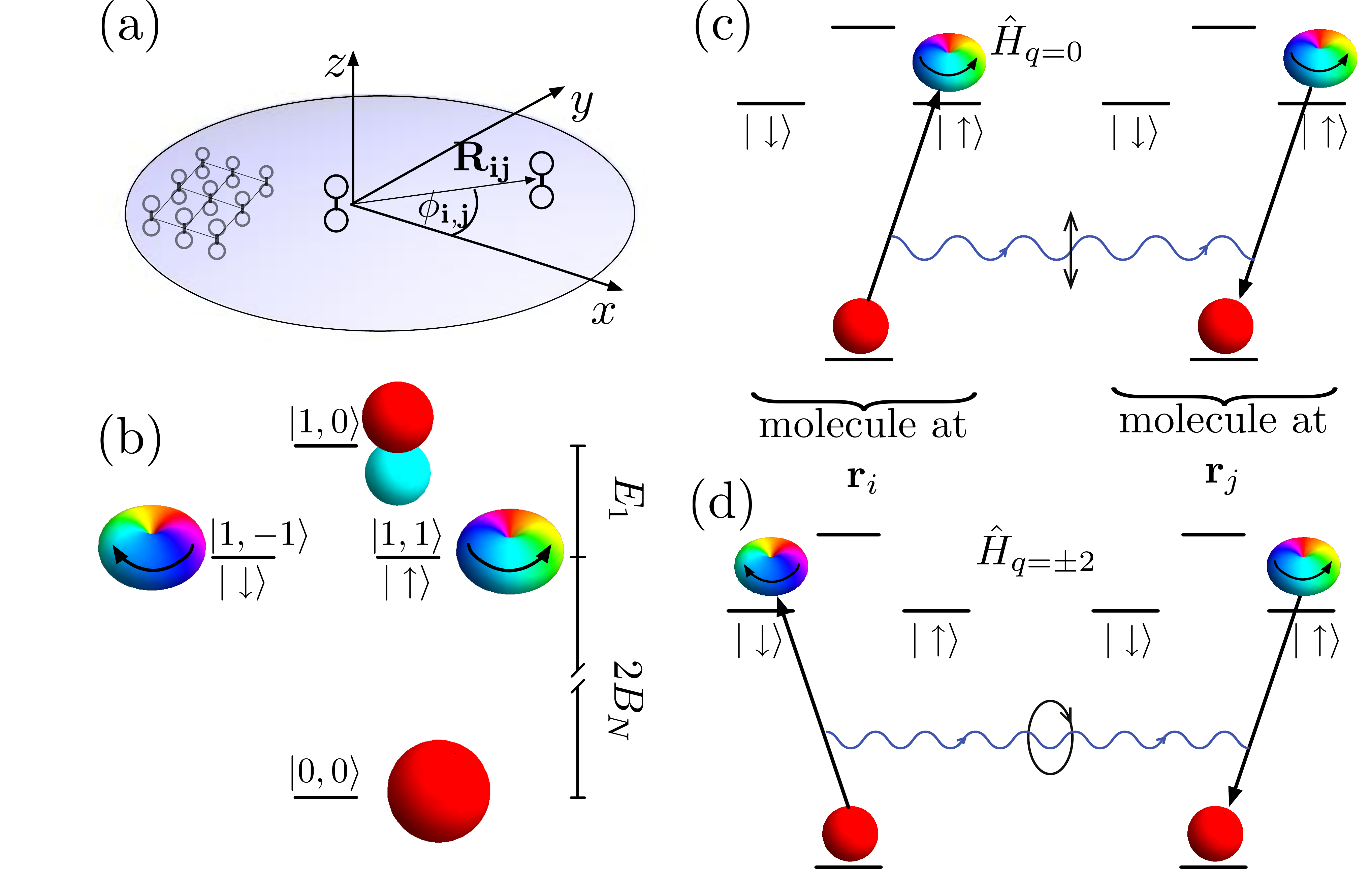}
	\caption{\label{fig:Schematic} {\bf The setup.} (a) Geometry: molecules confined  in the $xy$ plane,
	perpendicular to the quantisation axis $z$; $(R_{ij};\phi_{ij})$
	are the polar coordinates of the vector	$\mathbf{R}_{ij}$ joining molecules $i$ and $j$.
	(b) Rotational levels of a molecule.  The small icons mimic
	the angular distributions of the rotational states and are coloured according to their phase, the arrows showing
	the direction of the phase winding in the $|1,\pm1\rangle$ states.
	(c)-(d) Processes involving dipolar exchange interactions.
	In (c) molecules exchange rotational levels while preserving both their total internal angular
	momentum and its $z$-component.
	In (d) the internal angular momentum decreases by 2, while the orbital momentum of the two molecules
	increases by 2.
	}
\end{figure}

We consider an ensemble  of polar molecules confined
in a plane perpendicular to an external electromagnetic field that sets the quantisation axis $z$,
Fig.~\ref{fig:Schematic}(a). The rotational spectrum of each molecule can be indexed by the rotational
angular momentum $N$ and its projection $M$ onto the $z$ axis, Fig.~\ref{fig:Schematic}(b).
Throughout the paper we set $\hbar=k_{\mathrm{B}}=1$, and measure all lengths in units of the lattice constant $a$,
unless specified otherwise.

We assume that most molecules are in the ground rotational state ($N=0$),
and only the lowest-energy states,
those with $N=0$ and $N=1$, participate in the dynamics. The $N=1$ states are separated
from $N=0$ by a large gap $2B_N$ ($\sim$ GHz) which significantly exceeds the characteristic
interaction energy $E_{\mathrm{d}}$ ($\sim$ kHz). In addition,
the $\ket{1,0}$ state is separated from the $\ket{1,\pm1}$ states
by an energy scale $E_1$, $E_{\mathrm{d}}\ll E_1\ll B_N$, e.g., due to the presence of external electric field ${\bf E}$;
more details on realising such level structure will be given below.
Large lifetimes of the $N=1$ states ($\gtrsim10\,$s) allow us to neglect relaxation between the $N=1$ and $N=0$
manifolds.

The Hamiltonian for the system of polar molecules can be written as
\begin{align}
\hat{H}&=\sum_{i}\hat{H}_0\left(\mathbf{r}_i\right)+\frac{1}{2}\sum_{i\ne j}\hat{H}_{\mathrm{dip}}\left(\hat{\mathbf{d}}_i,\hat{\mathbf{d}}_j,\mathbf{r}_i-\mathbf{r}_j\right)\, ,
\end{align}
where the sums run over all particles in the system and $\hat{H}_0\left(\mathbf{r}_i\right)=\hat{\mathbf{p}}_i^2/2m+U(\mathbf{r}_i)+\hat{H}_{\mathrm{rot}}$ is the single-particle Hamiltonian.  Here, $m$ is the mass of a molecule,
$U(\mathbf{r})$ is the external periodic potential, and $\hat{H}_{\mathrm{rot}}=B_N\hat{\bN}^2-E_1 \hat{N}_z^2$ is the Hamiltonian of the internal degrees of freedom giving the spectrum in Fig.~\ref{fig:Schematic}(b).  The dipole-dipole interaction between two molecules $i$ and $j$ with dipole moments
$\hat{\bd}_i$ and $\hat{\bd}_j$ and separated by vector $\bR_{ij}$, Fig.~\ref{fig:Schematic}(a), is given by
\begin{align}
	 \hat{H}_{\mathrm{dip}}(\hat{\bd}_i,\hat{\bd}_j,\bR_{ij})=\frac{\hat{\bd}_i\cdot \hat{\bd}_j}{R_{i,j}^3}-3\frac{(\hat{\bd}_i\cdot \bR_{ij})(\hat{\bd}_j\cdot \bR_{ij})}{R_{i,j}^5},
	 \label{DipHam}
\end{align}
Introducing polar coordinates $\mathbf{R}_{i,j}=({R}_{i,j},\phi_{ij})$, the interaction Hamiltonian \eqref{DipHam}
can be decomposed as
\begin{align}
	&\hat{H}_{\mathrm{dip}}(\hat{\mathbf{d}}_i,\hat{\mathbf{d}}_j,\bR_{ij})=\hat{H}^{q=0}_{ij}+
\hat{H}_{ij}^{q=\pm2},
	\\
	&\hat{H}^{q=0}_{ij}=
	\frac{1}{R_{ij}^3}\left(\frac{\hat{d}_i^{-1}\hat{d}_j^{1}+\hat{d}_i^{1}\hat{d}_j^{-1}}{2}+{\hat{d}_i^z \hat{d}_j^z}\right),
	\label{eq:q=0}
	\\ &\hat{H}^{q=\pm2}_{ij}=-\frac{3}{2}\frac{1}{R_{ij}^3}\left(\hat{d}_i^{1}\hat{d}_j^{1}e^{-2i\phi_{ij}}
	+\hat{d}_i^{-1}\hat{d}_j^{-1}e^{2i\phi_{ij}}\right),
	\label{eq:q=pm2}
\end{align} where $\hat{d}^{\pm1}_i=\mp(\hat{d}^x_i\pm id_i^y)/\sqrt{2}$ are spherical components of the dipole operator of molecule $i$.

The operator $\hat{H}^{q=0}_{ij}$, Eq.~(\ref{eq:q=0}), conserves both the total
internal angular momentum of the interacting molecule and its $z$-component.
As shown in Fig.~\ref{fig:Schematic}(c), $\hat{H}^{q=0}_{ij}$
exchanges the states $|1,\pm1\rangle$ and $|0,0\rangle$ of two molecules while preserving $M_i+M_j$.  Such ``spin-exchange'' dipolar interactions have been  observed in recent
experiments on polar molecules \cite{Yan_Moses}, Rydberg atoms\cite{Weidemueller,PhysRevA.70.042703}, and magnetic atoms  \cite{Laburthe}.
 In contrast, the operator $\hat{H}^{q=\pm2}_{ij} $ transfers angular momentum between the internal and external orbital motion of the molecules, while preserving the total projection onto the $z$ axis.
 Namely, the operator $\hat{d}_i^{-1}\hat{d}_j^{-1}$
 decreases the internal angular momentum by $2$, while $e^{2i\phi_{ij}}$ increases the orbital angular momentum
 of a molecule by 2, thus preserving the total angular momentum, cf. Fig.~\ref{fig:Schematic}(d).  As we shall show, this transfer of angular momentum is responsible for the generation of the effective SOC of the elementary excitations.

\section*{Phenomenological analysis}
\label{PhenomAnal}

Let us assume that almost all the molecules are initially in their lowest rotational level $|0,0\rangle$
and are in a spatially uniform, not necessarily equilibrium state, but with a relaxation time
 sufficiently long to be considered stationary.  We then suppose that this state is slightly perturbed
by a resonant microwave pulse which excites  a small number of molecules from $|0,0\rangle$ to $|1,\pm 1\rangle$.
In what follows we show that the density and angular momentum dynamics of the $|1,\pm1\rangle$ rotational levels after such excitation is equivalent to that of an ideal gas of spin $1/2$ chiral quasiparticles (chirons).

The emergence of the excitations can be phenomenologically understood as follows.
Due to the translational invariance of the Hamiltonian of the system, the (quasi)momentum $\bk$
(in the presence of an optical lattice),
with polar coordinates $(k,\phi_\bk)$, Fig.~\ref{fig:Dispersion}(a), is a
good quantum number.
In the long-wave limit $\bk\rightarrow0$,  there is a degeneracy
between the excitations carrying molecules in the rotational states $\ket{\uparrow}\equiv\ket{1,1}$
and $\ket{\downarrow}\equiv\ket{1,-1}$ due to the symmetry
with respect to inverting the dipole moment. Note that  the other excited rotational states
are separated from $\ket{\uparrow}$ and $\ket{\downarrow}$ by large energy gaps and do not participate in the dynamics.
This allows us to consider a reduced space $\{\ket{\uparrow},\ket{\downarrow}\}$
of rotational states.  Each of these states can be obtained from the other by acting with the operators $(\hat{{d}}^{\pm1})^2$.
Hence, the most general form of the excitation Hamiltonian in the reduced space reads
\begin{eqnarray}
	\hh(\bk)=
	\left(
	\begin{array}{cc}
	\xi_\bk & \alpha(k)(k^{-})^2 \\
	\alpha(k)(k^{+})^2 & \xi_\bk
	\end{array}
	\right),
	\label{SpinOrbitHam}
\end{eqnarray}
where $k^{\pm}=k_x\pm ik_y$, and
$\alpha(k)$ and $\xi_\bk$ are some functions of  $k$.

The Hamiltonian in Eq.~(\ref{SpinOrbitHam}) describes quasiparticles with a two-branch spectrum with energies
\begin{eqnarray}
	E_\pm(\bk)=\xi_\bk\pm \alpha(k)k^2\, ,
	\label{ChironSpectra}
\end{eqnarray}
corresponding to the eigenstates
\begin{eqnarray}
	\psi_\pm(\bk)=\left(\pm e^{-i\phi_\bk}\quad e^{i\phi_\bk}\right)^T/\sqrt{2}
\end{eqnarray}
respectively,
and a Berry phase of $2\pi$.  In a system with an inversion-symmetric Hamiltonian, $\hh(\bk)=\hh(-\bk)$, the Berry phase 
can be defined\cite{BPdefinition} modulo $4\pi$ as an integral ${\it \Phi}_{\mathrm{BP}}=
-2i\int_C\bra{{\it \Psi}_\bq^\pm}\bnabla_\bq\ket{{\it \Psi}_\bq^\pm}d\bq$
along a contour $C$ connecting two points $\bk$ and $-\bk$ in momentum space.
Thus, the Berry phase $2\pi$ of chirons is {\it non-trivial}.
As is necessary in a system with time-reversal symmetry\cite{BlountBook,HaldaneBP}, ${\it \Phi}_{\mathrm{BP}}$ is a multiple of $\pi$.

Let us notice that if $\alpha(k)=const$ the Hamiltonian~(\ref{SpinOrbitHam})
coincides with that of the low-energy excitations in bilayer graphene~\cite{McCannFalko:bilayer,Novoselov}
in the wavelength limit $\bk\rightarrow0$.
In the next section we demonstrate that such Hamiltonian is indeed realised in the case of weak interactions
between the molecules.

\section*{Microscopic calculation of the Hamiltonian}

In the previous section we have shown phenomenologically that the effective Hamiltonian of long-wave
excitations in a system of polar molecules has the form given by Eq.~(\ref{SpinOrbitHam}). In what immediately
follows, we demonstrate that the Hamiltonian can be explicitly evaluated microscopically in the two opposite limits:
when the kinetic energy is negligible compared to the characteristic interaction strength
and when the interactions are small compared to the kinetic energy. While the  first regime can be achieved by pinning the   molecules in a deep optical lattice and is thus  relevant for  current experiments with reactive molecules \cite{Yan_Moses}, the second regime  could be in principle  realised in the future with molecules which are non-reactive and less susceptible to {\"u}berresonant processes~\cite{Mayle_13}.

To address the first limit,
we consider molecules in a deep, unit-filled square optical lattice. In this setting, the translational degrees of freedom are frozen and  dynamics occurs only  in the internal degrees of freedom.  The dynamics can be mapped  to that of a gas of bosons with spin $\sigma\in\left\{\uparrow,\downarrow\right\}$ and long-range hopping. The vacuum corresponds to all molecules being in the $|0,0\rangle$ state.  The {effective} Hamiltonian describing the rotational excitations can be expressed in terms of the bosonic creation $\hat{b}_{\mathbf{i}\sigma}^{\dagger}$ and annihilation $\hat{b}_{\mathbf{i}\sigma}$ operators of rotational excitations with $\sigma$ character at lattice site $\mathbf{i}=\{i_x,i_y\}$, as
\begin{align}
\nonumber  \hat{H}&=-{J_0}\sum_{\mathbf{i}\ne \mathbf{j};\sigma}\frac{\hat{b}_{\mathbf{i}\sigma}^{\dagger}\hat{b}_{\mathbf{j}\sigma}}{ |\br_\mathbf{i}-\br_\mathbf{j}|^3}-{J_2}\sum_{\mathbf{i}\ne \mathbf{j}}\frac{e^{-2i\phi_{ij}}\hat{b}_{\mathbf{i}\uparrow}^{\dagger}\hat{b}_{\mathbf{j}\downarrow}+
e^{2i\phi_{ij}}\hat{b}_{\mathbf{i}\downarrow}^{\dagger}\hat{b}_{\mathbf{j}\uparrow}}{|\br_\mathbf{i}-\br_\mathbf{j}|^3}\\
&\label{eq:HCHami}
\end{align} Here,
the hopping constants $J_0$ and $J_2$ are determined by dipole matrix elements.  We work in the hard core limit, which restricts the occupation number on each site to $0$ or $1$.  The hard-core constraint encapsulates that there is  at most one molecule per lattice site and each molecule can harbour at most one $N=1$ rotational excitation.  Physically, the hard-core constraint can stem either from strong elastic interactions or rapid inelastic loss rates, e.~g. two-body chemical losses, at short range~\cite{Yan_Moses,Zeno}.

\begin{figure}[t]
	\centering
	\includegraphics[width=0.55\columnwidth]{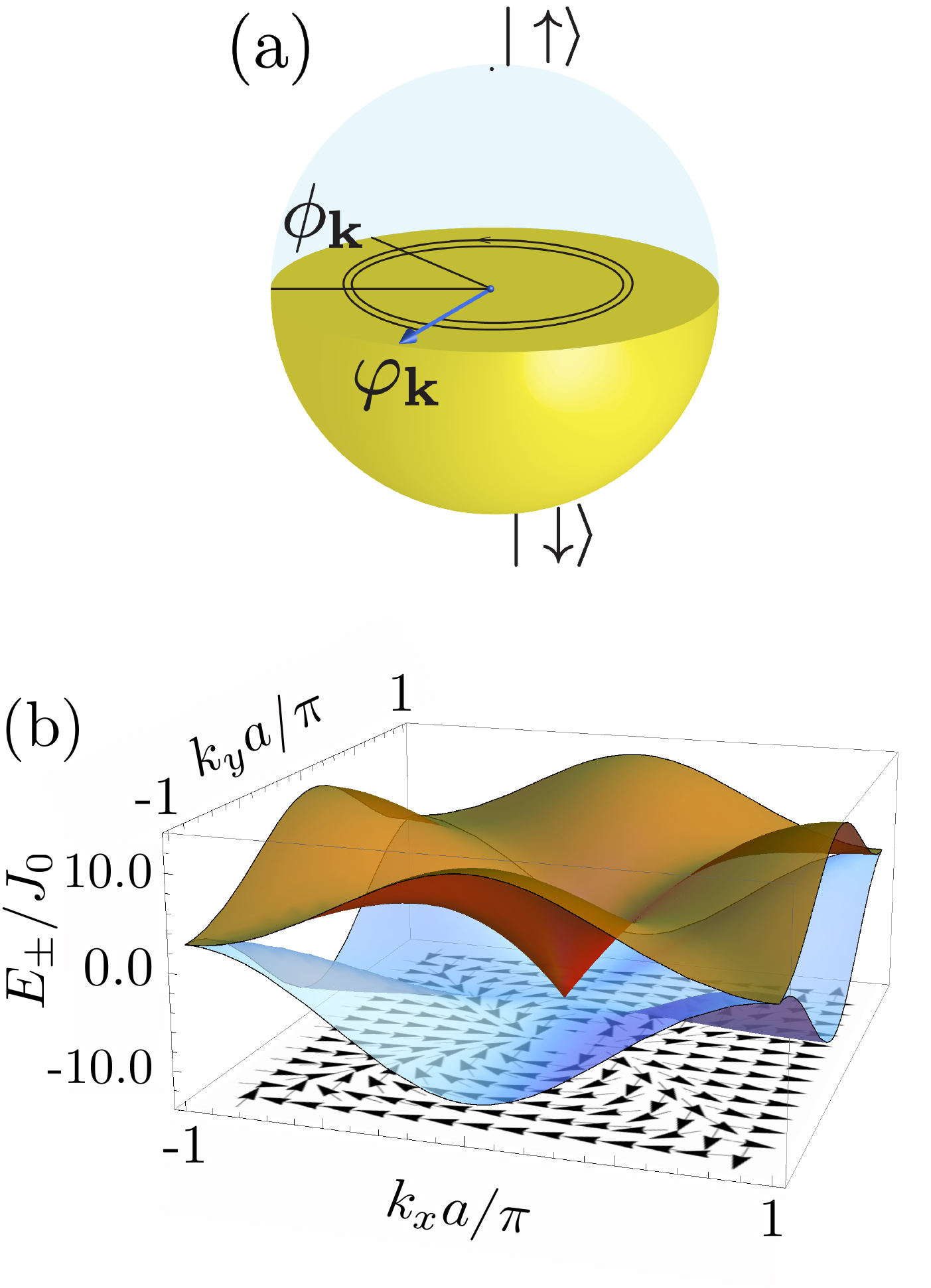}
	\caption{\label{fig:Dispersion} {\bf Chiron states.}  (a) Chiron state on the Bloch sphere.
	(b) Chiron dispersion in a deep optical lattice
	in the first Brillouin zone for $J_2=3J_0$. The lower (upper) surface corresponds to $E_+(\bk)$ [$E_-(\bk)$].
	The base of the plot shows the orientation of the $E_+$ branch Bloch vector in the equatorial plane.
	}
\end{figure}

The dispersions of a single rotational excitation are given by
\begin{align}
\label{eq:lattenergy}E_{\pm}\left(\mathbf{k}\right)&=-J_0F^{(0)}\left(\mathbf{k}\right) \mp {J_2} |F^{(2)}\left(\mathbf{k}\right)|,
\end{align} and  shown in Fig.~\ref{fig:Dispersion}.  Here, $F^{(n)}(\mathbf{k})=\sum_{\mathbf{j}\ne 0}\exp(-i \mathbf{k}\cdot \br_\mathbf{j}+ i n\phi_{\mathbf{j}})|\br_\mathbf{j}|^{-3}$ with $\br_\mathbf{j}$ a vector connecting sites in the square lattice and $\phi_{\mathbf{j}}$ the polar angle of $\br_\mathbf{j}$.
The phase $\varphi_{\mathbf{k}}$ of  $F^{(2)}$, {\it i.e.} $F^{(2)}(\mathbf{k})=|F^{(2)}\left(\mathbf{k}\right)| e^{i \varphi_{\mathbf{k}}}$ determines the polar angle of the Bloch vector, Fig.~\ref{fig:Dispersion}(a).
In the long-wave limit, $\bk\rightarrow0$,  we obtain in accordance with Eq.~\eqref{SpinOrbitHam} that
 $\varphi_{\mathbf{k}}\approx2\phi_{\mathbf{k}}$, $\xi_{\mathbf{k}}/J_0\approx  A +2\pi/k$ and $\alpha\left(k\right)=2\pi J_2/(3k)$, with $A\approx 9.03$.  For general ratios $J_2/J_0$, both branches have a conical dispersion for small $k$.  For the case $J_2=3J_0$, as results from the geometry of Fig.~\ref{fig:Schematic}(a), there is a cancellation of the linear $k$ component in the $E_{+}$ branch.  This leads  to  a locally flat  dispersion $E_{+}(\mathbf{k})/J_0\approx A+\mathcal{O}(k^2)$ and a conical dispersion $E_{-}(\mathbf{k})/J_0\approx A +4\pi k+\mathcal{O}(k^2)$, Fig.~\ref{fig:Dispersion}(b).

\begin{figure*}[t]
	\centering
	\includegraphics[width=1.8\columnwidth]{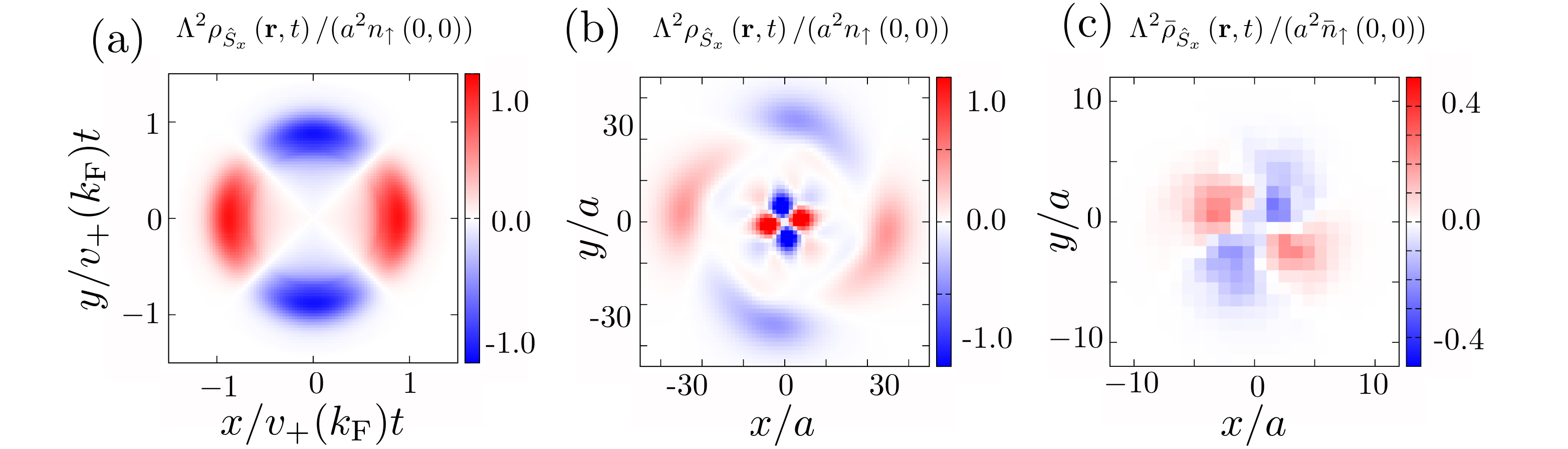}
	\caption{\label{fig:ChironDynam} {\bf Chiron spin dynamics. } Spatial distributions of the spin coherences in a gas of polar
	molecules after applying a focused resonant pulse.
	(a) Density profile of $\hat{S}_x$ in a weakly-interacting Fermi liquid.
	(b) Density profile of $\hat{S}_x$ in a deep optical lattice.
	(c)  The density of $\hat{S}_x$, analogous to (b) but  for a lattice of 10\% filling, averaged over disorder realisations (overbar denotes disorder average).  The peak magnitude of $\bar{\rho}_{\hat{S}_x}$ is reduced by a factor of $\sim 20$ compared to unit filling, but the symmetry is the same.	}
\end{figure*}

In the case of {\it a sufficiently shallow optical lattice} or {\it weak interactions},
the kinetic energy of the molecules can dominate over the mean interaction energy.
In this case, the dynamics can be analysed perturbatively in the interactions.
By explicitly evaluating the Hamiltonian of the excitations, the details of which are provided in the Supplementary Methods,
we reproduce Eq.~(\ref{SpinOrbitHam}) with the off-diagonal entry
\begin{eqnarray}
	\hh_{\downarrow\uparrow}(\bk)=\pm\frac{\pi|\bd|^2}{6}\int_\bq\:  \frac{(q^+)^2}{q} f_{00}(\bq+\bk),
	\label{AlphaPert}
\end{eqnarray}
where the upper and the lower signs apply to bosonic and fermionic molecules respectively;
$f_{00}(\bq)$ is the distribution function of the molecules in the $\ket{0,0}$ state. It is
assumed to be stationary and independent of the molecule position but it is not restricted to be  in thermal equilibrium;
$q^+=q_x+ iq_y$, and $\int_\bq\ldots=\int(2\pi)^{-2}\ldots d^2\bq$.  Due to the smallness of the interactions, the diagonal elements of the matrix in  Eq. (\ref{SpinOrbitHam})
are close to the kinetic energy of a single molecule and are only slightly modified by the interactions,
$\xi_\bk\approx k^2(2m)^{-1}\left[1+\mathcal{O}(|\bd|^2)\right]$.  This calculation is performed explicitly in the Supplementary Methods.

The long-time dynamics is dominated by small momenta. For an isotropic distribution
function $f_{00}(\bq)=f_{00}(q)$,
from Eq.~(\ref{AlphaPert}) we find the value of the spin-orbital coupling in the limit $\bk\rightarrow0$ to be
\begin{eqnarray}
\label{eq:alphaIso}	\alpha=\pm\frac{|\bd|^2}{32}\int_0^{\infty}
	f_{00}(q) \:dq.
\end{eqnarray}

For a Fermi liquid of fermionic molecules at zero temperature [$f_{00}(q)=\theta(k_{\mathrm{F}}-q)$] Eq.~\eqref{eq:alphaIso} yields
\begin{equation}
	\alpha=-|\bd|^2 k_{\mathrm{F}}/32,
\end{equation}
where $k_{\mathrm{F}}$ is the Fermi momentum. At sufficiently high temperatures $T\gg n/m$, where $n$ is the density of the molecules
(per nuclear spin), the distribution function is close to that of a Boltzmann gas,
$f_{00}(q)\approx\frac{2\pi n}{mT}e^{-\frac{q^2}{2mT}}$, and
\begin{equation}
	\alpha=\pm\frac{\pi\sqrt{\pi}|\bd|^2 n}{ 16\sqrt{ 2 m T}}.
\end{equation}
For cold atoms in a quadratic trapping potential, the density of the molecules depends on temperature
as $n(T)\propto T^{-1}$, resulting in the temperature dependency of the SOC
$\alpha(T)\propto T^{-3/2}$.

\section*{Chirality manifestations : Spin and density dynamics}

The chirality of the excitations can be observed in the dynamics of the spin-1/2  operator, $\bS=\{\hat{S}_x,{\hat S}_y,{\hat S}_z\}$,  in the reduced space of the rotational levels $\ket{\uparrow}$ and $\ket{\downarrow}$.  Let us assume that a short laser pulse excites
a group of molecules in a small region of characteristic
size $\Lambda$ around $\br=0$, $\ket{0,0}\rightarrow\ket{\uparrow}$ 
(the results of this paper can be easily generalised to include more general
excitation protocols, $\ket{0,0}\rightarrow A_{\uparrow}\ket{\uparrow }+A_{\downarrow}\ket{\downarrow}$).  The internal state $|\uparrow\rangle$ corresponds to excitation by light with right-circular polarisation $x+iy$, and has a definite phase winding, as shown in Fig.~\ref{fig:Schematic}(b).  Hence, for a spatially isotropic distribution of excitations, the laser polarisation is what determines the spatial phase pattern emerging during the dynamics.

For sufficiently small $\Lambda$ chirons leave the excited region quickly, reaching sufficiently low density,
so that interactions between them can be neglected,
and their dynamics is described by the kinetic equation for free particles, see Methods.
In principle, chiron-chiron interactions may be important for hard-core particles in a deep optical lattice
in the beginning of the dynamics, which, however, will not affect the results qualitatively.
In the limit of pinned molecules
we additionally check that chiron-chiron interactions can be neglected by numerically
simulating the dynamics of two excitations, see the Supplementary Methods and Supplementary Figures 2,3.

\begin{figure*}[t]
	\centering
	\includegraphics[width=1.5\columnwidth]{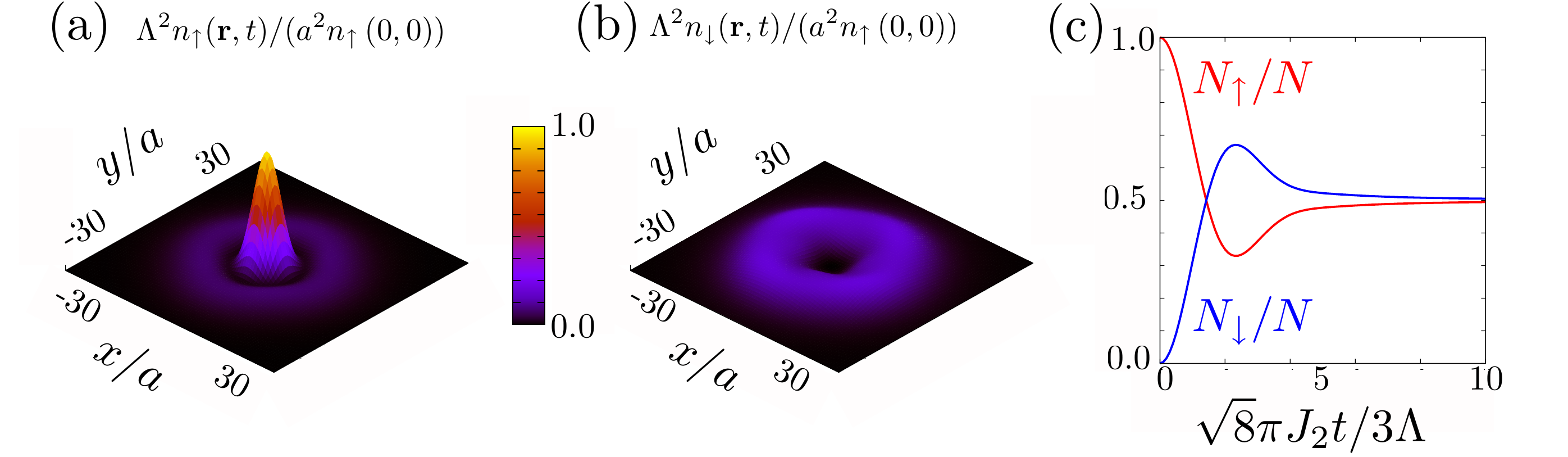}
	\caption{\label{fig:DenDynamics}  {\bf Chiron density dynamics. } Densities of $|\uparrow\rangle$ (panel (a)) and $|\downarrow\rangle$ (panel (b)).  The vortex structure is clearly visible in the $|\downarrow\rangle$ density.  The density dip $n_{\downarrow}\left(\mathbf{r},t\right)\propto r^4$ in the centre of the vortex structure is a manifestation of the Berry phase $2\pi$.  (c) Dynamics of the total spin populations in a deep optical lattice. The population of the two spin components undergo a single oscillation before monotonically approaching 1/2. }
\end{figure*}

The chiral nature of the excitations is clearly visible in the density of $\hat{S}_x$
\begin{equation}
	\rho_{\hS_x}(\br,t)=
	\frac{1}{2}\sum_\pm
	\pm\int_\bq
	 f_{\uparrow\uparrow}\left[\br-t\bv_\pm(\bq),\bq\right]
	\cos(2\phi_\bq),
	\label{rhoS}
\end{equation}
where $f_{\uparrow\uparrow}(\br,\bq)$ is the distribution function
of the molecules in the $\ket{\uparrow}$ state at time $t=0$, $\bv_\pm(\bq)$ are the velocities of the
two chiron branches, and $\phi_{\bq}$ is the polar angle of the vector $\bq$.
In the long-wave limit (see Eq.~(\ref{ChironSpectra}))
\begin{equation}
	\bv_\pm(\bq)=\bnabla_\bq\xi\pm\left(2\alpha(q)\bq+q^2\bnabla_\bq\alpha(q) \right).
\end{equation}
Eq.~(\ref{rhoS}) describes the spin distribution at sufficiently long times,
when it is dominated by long-wave
chirons ($q\ll a^{-1}$). In this limit the  phase factor $\cos(2\phi_\bq)$ originates from the off-diagonal element of  Eq.  \ref{SpinOrbitHam}. To account for arbitrary-momenta excitations, $2\phi_\bq$ in Eq.~(\ref{rhoS})
has to be replaced by the polar angle $\varphi_\bq$ of a chiron state on the Bloch sphere, Fig.~\ref{fig:Dispersion}(a).  For the small $\Lambda$ under consideration, the distribution has a $d$-wave symmetry
$\rho_{\hS_x}\propto \cos(2\phi)$ at long times, Figs.~\ref{fig:ChironDynam}(a-b),
which is a manifestation of the non-trivial Berry phase $2\pi$ of the excitations.  The radial distribution of the spin component $\rho_{\hS_x}$ after applying a narrow laser
pulse depends on the molecular statistics, interaction strength, optical lattice depth, etc., while its
$d$-wave symmetry is universal, being a consequence of the
Berry phase $2\pi$.

The spatial distribution of the spin coherences can be particularly easily understood in the case
of a Fermi liquid (fermionic molecules at low temperatures) with weak interactions.
In this case, the excitations propagate at the maximal speed $v_+(k_{\mathrm{F}})$, $k_{\mathrm{F}}$ being the Fermi momentum.
At a distance $r$ away from the initial excitation pulse, the spin distribution remains unaltered
until time $t=r/v_+(k_{\mathrm{F}})$ when it is reached by the quickest branch of chirons with the
angular distribution of spins $\rho_{{\hat S}_x}\propto\cos(2\phi)$.
At a slightly later moment of
time $t=r/v_-(k_{\mathrm{F}})$ the same point is reached by a wave of slower chirons with opposite spin,
after which the spin density remains very small. The resulting distribution
is shown in Fig.~\ref{fig:ChironDynam}(a).

In the case of a deep optical lattice, the branch of chirons with the dispersion $E_-(\bq)$
is significantly faster than the other branch, leading to a quick spatial
separation of the two branches, Fig.~\ref{fig:ChironDynam}(b), after applying the laser pulse.
The outer circular density front corresponds to the faster chirons, which propagate at a nearly
constant speed, while the more complex
inner pattern comes from the slower branch of chirons. Despite different dispersions of the chirons,
the angular distribution of the spin is again $\propto\cos(2\phi)$.


Because of the SOC spin is not conserved
and the total numbers of molecules
in the rotational states $\ket{\uparrow}$ and $\ket{\downarrow}$,
{\small\begin{align}
	 &N_{\uparrow}(t)=
	\int d\br\int_\bq f_{\uparrow\uparrow}(\br,\bq)\cos^2[{(E_+(\bq)-E_-(\bq))t}/{2}],
	\\
	 &N_{\downarrow}(t)=
	\int d\br\int_\bq f_{\uparrow\uparrow}(\br,\bq)\sin^2[{(E_+(\bq)-E_-(\bq))t}/{2}],
\end{align}}
are time-dependent.
At long times $t\rightarrow\infty$,
both $N_\uparrow(t)$ and $N_\downarrow(t)$
saturate at a half of the number $N_{\mathrm{ex}}$ of the initially excited molecules,
Fig.~\ref{fig:DenDynamics}(c), regardless of the details of the
distribution function $f_{\uparrow\uparrow}(\br,\bq)$.
The total number of molecules in excited rotational states is conserved, $N_\uparrow(t)+N_\downarrow(t)=N_{\mathrm{ex}}$.

Thus, the SOC transfers the internal angular momentum of the molecules, all of which are in the
$\ket{\uparrow}$ state at $t=0$, to their orbital motion, leading to the formation of a
{\it vortex structure} around $\br=0$. This manifests itself, for instance, in a dip in the density of
molecules in the $\ket{\downarrow}$ rotational state, $n_\downarrow(\br)\propto r^{2{\it \Phi}_{\mathrm{BP}}/\pi}=r^4$,
around the centre of the vortex structure, Fig.~\ref{fig:DenDynamics}(b).  In particular, if the chiron spectrum
has a branch with quadratic dispersion $E(\bk)\sim k^2(2M)^{-1}$,
which is realised, for example, for weak interactions in a shallow optical lattice or
for strong interactions in a deep optical lattice, the density of the spin-down molecules close
to the centre of the vortex estimates $n_\downarrow(\br,t)\sim\left(M^2r^2/t^2\right)^2N_{\mathrm{ex}}/\Lambda^2$
at sufficiently long times.  Far from the centre of the vortex the density profile is described by freely propagating chirons which
do not interfere with each other:
\begin{eqnarray}
	n_{\uparrow}(\br,t)= n_{\downarrow}(\br,t)=	\frac{1}{4}\sum_\pm\int_\bq
	f_{\uparrow\uparrow}\left[\br-t\bv_\pm(\bq),\bq\right].
	\label{density}
\end{eqnarray}
The two contributions in the sum in Eq.~(\ref{density}) correspond to the two branches of chirons
propagating with velocities $v_+(\bq)$ and $v_-(\bq)$, which leads to their spatial separation.

\section*{Experimental Accessibility}
\label{ExpAcc}

In this section, we discuss some details related to the observation of dipolar SOC in present cold polar molecule experiments, taking as a representative example the KRb experiment at JILA~\cite{Yan_Moses}. To prevent chemical reactions, KRb polar molecules are pinned in a deep 3D optical lattice.  The relevant energy scales for Eq.~\eqref{eq:HCHami} are $J_0=|\langle 1,1|\hat{d}^1|0,0\rangle|^2/(2a^3)\sim 100 h\,$Hz and $J_2=3J_0$.  Trapping in a deep optical lattice may also be required for molecular species which are chemically stable, as the presence of a very high density of resonances at ultracold energies has been proposed to lead to long-lived collision complexes which are highly susceptible to three-body loss~\cite{Mayle_13}.

The chirons' spectra in the entire Brillouin zone (BZ) can be measured by means of Rabi spectroscopy provided the probe beam can transfer the required quasi-momentum  $k_{\mathrm{R}} \simeq 1/a$ to the molecules.  Direct microwave transitions are insufficient since they have $k_{\mathrm{R}} a\ll 1$, but $k_{\mathrm{R}}a \simeq 1$ can be achieved using  optical Raman pulses.  Here, $k_{\mathrm{R}}=|\mathbf{k}_1-\mathbf{k}_2|$, with $\mathbf{k}_i$ the wavevector of the $i^{th}$ Raman beam.  Raman transitions between internal states are already a key part of the production of ground-state molecules through STIRAP~\cite{KRb}; our proposal requires only minor modifications of this well-established procedure.

Due to the inherent difficulty of directly cooling molecules, present experiments are not quantum degenerate, leading to a sparse lattice filling  fraction near 10\%~\cite{Yan_Moses,Zeno,Hazzard_Gadway_14}.  Fig.~\ref{fig:ChironDynam}(c) displays the density of $\hat{S}_x$ in a lattice with 10\% filling, averaged over disorder realisations.  The $d$-wave symmetry of the distribution is still visible, albeit with reduced contrast compared to the unit-filled case (Fig.~\ref{fig:ChironDynam}(b)).  The $d$-wave symmetry is a consequence of the Berry phase, a topological property, and so is robust against disorder. In contrast, disorder smears the vortex structure.

In the rotational structure of KRb, nuclear quadrupole interactions cause the states with predominant $|1,-1\rangle$ and $|1,1\rangle$ character to be non-degenerate by about 70$h$kHz at the $545$G magnetic fields used for magneto-association~\cite{AnisotropicPol}.  These states can be made degenerate  and out of resonance with the $|1,0\rangle$ level by increasing the strength of the magnetic field $B$, even at zero electric field. In this scenario, the $B$ field determines the quantisation axis.  For $^{40}$K$^{87}$Rb, where the nuclear quadrupole moments are $(eqQ)_{\mathrm{Rb}}=-1.308h\,$MHz and $(eqQ)_{\mathrm{K}}=0.452h\, $MHz~\cite{AnisotropicPol}, the levels cross near $B\simeq $1260G, well within experimental feasibility.  The energy difference between these levels close to the crossing is nearly linear, with a slope of roughly 40Hz$\,$G$^{-1}$.  Stabilisation of magnetic fields at the 10mG level, which is routine in ultracold gas experiments, would correspond to non-degeneracy on the order of 0.4Hz, and will not significantly affect our results.  Similar comments apply for the other alkali metal dimers.  The level structure in Fig.~\ref{fig:Schematic}(b) also results for $^1\Sigma$ molecules without hyperfine structure, for example bosonic SrO, in the presence of a uniform electric field.

Finally, we note that several knobs  can be used to manipulate the Hamiltonian, Eq.~\eqref{eq:HCHami}.  For example, by changing the angle of the quantisation axis with respect to the space-fixed $z$ axis one can tune the ratio $J_2/J_0$ and remove the cancelation seen in the $E_+$ branch. This can be used in turn to control  the propagation velocity of the two branches of chirons.   Chiron-chiron interactions which fall off as $1/r^3$ can also be controllably introduced by turning on an external static electric field.

\section*{Discussion}

We have demonstrated that
a 2D system of polar molecules behaves as a gas of chiral excitations with a  Berry phase $2\pi$. We have shown that signatures of those excitations,  which resemble the low-energy excitations exhibited by bilayer graphene, manifest in both the dynamics of the
density and spin coherences.

The implementation of SOC in polar molecules presented here can open other exciting research
avenues. In particular, by superimposing an effective magnetic field to the chirons via  light-generated synthetic gauge fields \cite{Galitski2013}  it might be possible to simulate  the unconventional quantum Hall effect of bilayer graphene and to see its intriguing consequences \cite{McCannFalko:bilayer}  in the lowest Landau levels in the limit of low
chiron density  and high synthetic  magnetic fields.  Such an effect would require fermionic statistics of the excitations, which can be realised with fermionic molecules in a shallow optical lattice.

Although so far  the system in consideration  is an ideal or nearly ideal gas of chirons and interactions between them can be neglected, chiron-chiron interactions tunable by
 the duration of the laser pulse, the size of the excited region or  an external electric field
may lead to very rich physics. For example, chiron-chiron interactions together with non-stationary background $N=0$ molecules  or microwave dressing can give rise to interesting dynamic structures, new types of transport phenomena and even to
 fractional quantum Hall phases when combined  with synthetic gauge fields \cite{Yao2013,PhysRevLett.110.185301}.

Finally, we expect our predictions to be extendable to
other dipole-dipole interacting systems such as Rydberg
atoms, magnetic atoms, and magnetic defects
in solids.  \Notes{}

{\it Acknowledgements.} We appreciate useful discussions with M.~Hermele, M. Lukin, N. Yao, K.R.A.~Hazzard, T. Pfau, and the KRb JILA experimental group.
This work has been financially supported by
 NIST, JILA-NSF-
PFC-1125844, NSF-PIF-1211914, NSF-PHY11-25915, ARO, ARO-DARPA-OLE,
AFOSR, AFOSR-MURI,  and the NSF grants DMR-1001240 and PHY-1125844. M.L.W thanks the NRC postdoctoral
fellowship program for support. S.V.S. has been also partially supported by the Alexander
von Humboldt Foundation
through the Feodor Lynen Research Fellowship. A.M.R. and V.G.
thank the Aspen Center for Physics and  KITP.

{\it Author contributions.}
All authors contributed significantly to the work presented in this paper.

{\it Competing financial interests.}
The authors declare no competing financial interests.

\section*{Methods}

\subsection*{Kinetic equation for polar molecules}

To characterise the dynamics of a system of polar molecules, we introduce
the non-equilibrium Green's functions
\begin{eqnarray}
	G^<_{\sigma\sigma^\prime}(\br_1,t_1;\br_2,t_2)=
	\mp i\langle\hPsi^\dagger_{\sigma^\prime}(\br_2,t_2)\hPsi_\sigma(\br_1,t_1)\rangle,
	\label{Glesser}
	\\
	G^>_{\sigma\sigma^\prime}(\br_1,t_1;\br_2,t_2)=
	-i\langle\hPsi_\sigma(\br_1,t_1)\hPsi^\dagger_{\sigma^\prime}(\br_2,t_2)\rangle,
	\label{Glarger}
\end{eqnarray}
where $\sigma$ and $\sigma^\prime$ label the internal states of the molecules [$\sigma$ corresponds to $(|1,-1\rangle,|0,0\rangle,|1,1\rangle)$].
The upper (lower) sign applies to bosonic (fermionic) particles.

The distribution functions of the molecules are defined as
\begin{equation}
	f_{\sigma_1\sigma_2}(t,\br,\bp)=
	\pm(2\pi i)^{-1}\int G_{\sigma_1\sigma_2}^<(t,\br,\bp,E)dE,
	\label{f}
\end{equation}
where the upper and the lower signs apply to bosonic and fermionic particles respectively,
and $G_{\sigma_1\sigma_2}^<(t,\br,\bp,E)$ is the result of the
Wigner-transformation\cite{RammerSmith,Kamenev:book} of
$G^<_{\sigma\sigma^\prime}(\br_1,t_1;\br_2,t_2)$.

Using the functions (\ref{Glesser}) and (\ref{Glarger}),
we define a $2\times 2$ matrix in the Keldysh space\cite{Kamenev:book,RammerSmith},
\begin{eqnarray}
    \underline G=\left(
    \begin{array}{cc}
    G^{\mathrm{R}} & G^{\mathrm{K}} \\
    0 & G^{\mathrm{A}}
    \end{array}
    \right).
    \label{Gunderline}
\end{eqnarray}
Each of the Green's functions in Eq.~(\ref{Gunderline}) is a matrix in the space of the
internal rotational levels of the molecules.

The function (\ref{Gunderline}) satisfies the equation
\begin{eqnarray}\label{Dyson}
   [(\underline G_0^{-1}-\underline\Sigma) \otimes \underline
   G]=0
\end{eqnarray}
(Dyson equation minus its conjugate), where
$G_0^{-1}(1,2)=[i\partial_{t_1}-\hat H(\br_1)]\delta(1-2)$;
$1=(t_1,\br_1)$, $2=(t_{2},\br_{2})$, and $\underline\Sigma$
is the self-energy part, determined by the dipole-dipole interactions.

In terms of the distribution functions $f_{\sigma_1\sigma_2}$ the kinetic equation reads
\begin{align}
	&\partial_t f_{\sigma_1\sigma_2}-\bnabla U\:\partial_\bp f_{\sigma_1\sigma_2}
	\nonumber\\
	&+\partial_\bp(E_\bp\delta_{\sigma_1\sigma}+\Sigma_{\sigma_1\sigma})\bnabla f_{\sigma\sigma_2}
	=\left(\mathrm{St}f\right)_{\sigma_1\sigma_2},
	\label{kingenmol}
\end{align}
where $U(\br)$ is the external smooth (trapping) potential, $E_\bp$ is the kinetic
(quasi)energy, the summation over repeated indices is implied,
and $\mathrm{St}f$ is the collision integral, which accounts for the relaxation of the distribution
function due to molecular collisions.

In this paper we consider a model with a small relaxation rate, which can be neglected on the characteristic
times of interest, so that the excitations propagate ballistically. Also, we assume
that chirons reach sufficiently low density shortly after they are excited, so that chiron-chiron interactions
can be neglected, and the problem becomes effectively single-particle.

Introducing the chiron annihilation operator
\begin{equation}
	\hPsi{(\bk)}_\pm=\left[e^{i\phi_\bk}\hPsi_\downarrow{(\bk)}\pm e^{-i\phi_\bk}\hPsi_\uparrow{(\bk)}\right]/\sqrt{2},	
\end{equation}
where $\hPsi_{\uparrow,\downarrow}{(\bk)}$ are the annihilation operators for the plane-wave states in
the reduced space $\{\ket{\downarrow},\ket{\uparrow}\}$ of the rotational levels of the molecules, the kinetic
equation (\ref{kingenmol}) is reduced, under the assumptions made above, to that for a single-particle problem with
the dispersion $E_\pm(\bk)$:
\begin{align}
	&\partial_t f_{\pm}-\bnabla U\:\partial_\bp f_{\pm}
	+\bv_\pm(\bp)\bnabla f_{\pm}
	=0.
	\label{KineticChiron}
\end{align}

In the case $U=0$, considered in this paper, the most general solution of Eq.~(\ref{KineticChiron}) reads
\begin{equation}
	f_\pm(t,\br,\bp)=G_\pm[\br-t\bv_\pm(\bp),\bp],
\end{equation}
$G_\pm(\br,\bp)$ being arbitrary functions of two arguments.



\renewcommand{\thetable}{S\arabic{table}}

\renewcommand{\figurename}{Suplementary Figure}

\setcounter{equation}{0}
\setcounter{figure}{0}

\widetext

\begin{figure*}[t]
	\centering
	\includegraphics[width=0.4\columnwidth]{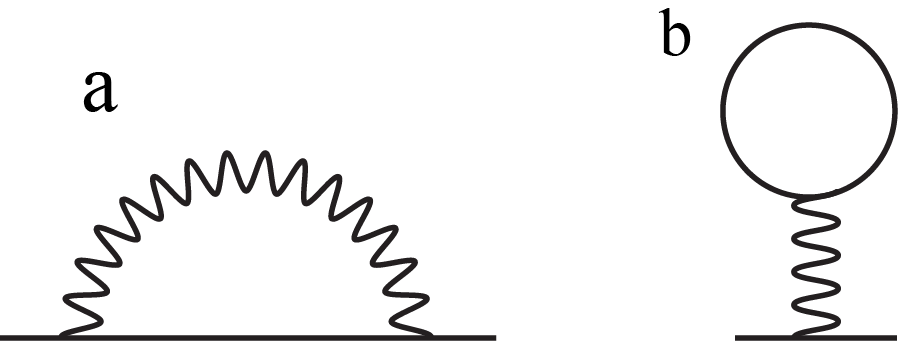}
	\caption{\label{SigmaLowest} Leading contributions to the self-energy part.}
\end{figure*}



\begin{figure*}[t]
	\centering
	\includegraphics[width=0.9\columnwidth]{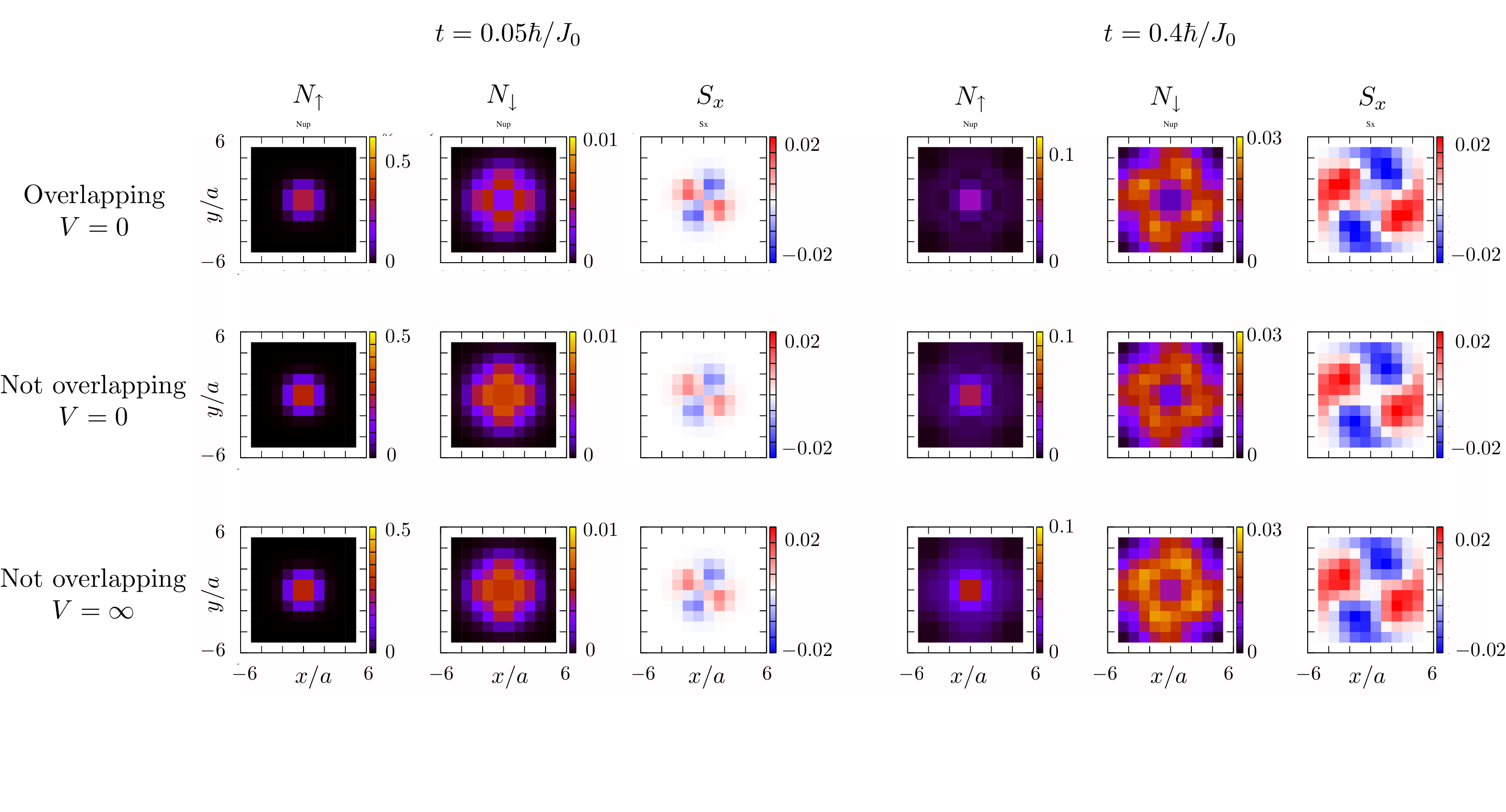}
	\caption{\label{fig:Supp1}  Dynamics of two rotational excitations with Gaussian character for: overlapping initial wavepackets and no interactions (top row), non-overlapping initial wavepackets and no interactions (middle row), and non-overlapping initial wavepackets and hard-core interactions (bottom row).}
\end{figure*}


\begin{figure*}[t]
	\centering
	\includegraphics[width=0.9\columnwidth]{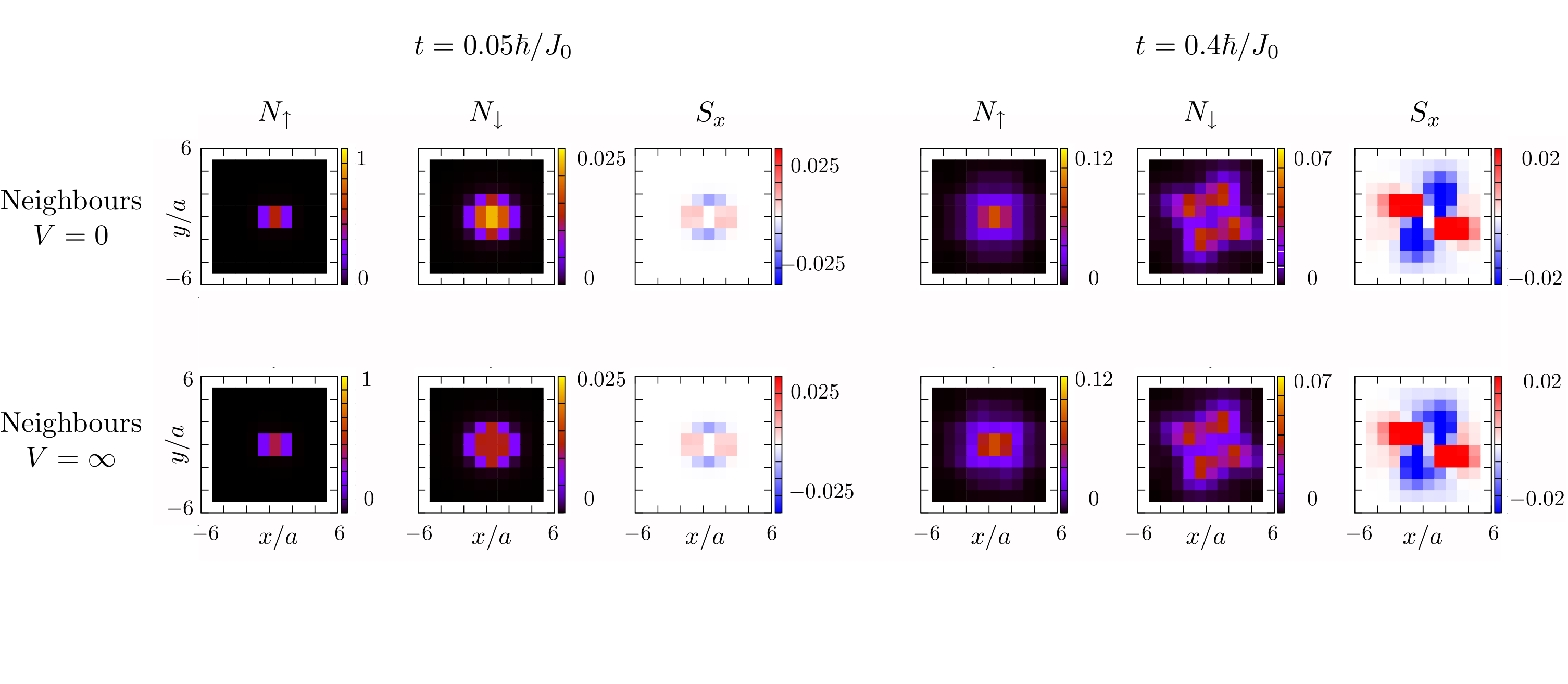}
	\caption{\label{fig:Supp2}  Dynamics of two rotational excitations localised at neighbouring lattice sites for: no interactions ($V=0$, top row) and hard-core interactions ($V=\infty$, bottom row).}
\end{figure*}

\clearpage

\section*{ Supplementary Methods}

\subsection*{Spin-orbital coupling from weak interactions}

In the limit of weak interactions the chirons' spectra can be obtained perturbatively in the
interaction strength. The leading-order contribution to the self-energy part $\underline\Sigma$
is shown in Figure~\ref{SigmaLowest}. The wiggly and solid lines correspond to the molecule and
interaction propagators respectively.

Because of the instantaneous character of the interactions, we can define an interaction propagator
which is diagonal in Keldysh space  \cite{KeldNote},
\begin{eqnarray}
    \underline D=\left(
    \begin{array}{cc}
    D & 0 \\
    0 & D
    \end{array}
    \right),
\end{eqnarray}
$D(\bq)$ being the Fourier-transform of the interaction Hamiltonian.
The interaction vertices
between the molecules and the interaction propagators correspond to the
respective matrix elements of the dipole operator and the
matrices $\gamma_{ij}^k$ and $\tilde\gamma_{ij}^k$ in Keldysh space\cite{RammerSmith} for
the absorption
and emission of interaction quanta, with the indices $ij$ and $k$
acting in the molecular and interaction Keldysh spaces respectively,
$\gamma_{ij}^1=\tilde\gamma_{ij}^2=\delta_{ij}/\sqrt{2}$,
$\gamma_{ij}^2=\tilde\gamma_{ij}^1=\tau_{ij}^1/\sqrt{2}$.
The only important dipole matrix elements for the generation of the spin-orbital coupling
are
\begin{eqnarray}
	\bra{1,\pm 1}\bd^{\pm 1}\ket{0,0}=|\bd|/\sqrt{3}.	
\end{eqnarray}

The process in Figure~\ref{SigmaLowest}(b) conserves the molecules' spin and, thus, does not lead
to the spin-orbital coupling.
This contribution results only in the renormalisation of the energy
gaps between the rotational levels $\ket{0,0}$ and $\ket{1,0}$
and between the levels $\ket{0,0}$ and $\ket{\uparrow,\downarrow}\equiv\ket{1,\pm 1}$.
The diagram in Figure~\ref{SigmaLowest}(a) allows  the incoming and
the outgoing states to exchange spins. It describes both the interactions-generated spin-orbital coupling
and a renormalisation of the molecules' kinetic energy.
\begin{align}
	\delta\xi_\bk=\frac{|\bd^2|}{24}\int_\bq F^{(0)}(\bq)f_{00}^\prime(q)
	\left[k^2-(\bq\cdot\bk)^2/q^2\right]/q
	+\frac{|\bd|^2}{24}\int_\bq F^{(0)}(\bq)f_{00}^{\prime\prime}(q)(\bq\cdot\bk)^2/q^2,
\end{align}
where the function $F^{(0)}(\bq)$, defined in the main text after Eq.~(9), is the Fourier-image of $1/r^3$
on the lattice, $f_{00}^\prime(q)$ and $f_{00}^{\prime\prime}(q)$ are respectively the first and the second derivatives
of the distribution function $f_{00}(q)$. Such renormalisation
is equivalent to a small modification of the molecules' mass, so long as the interactions
are small, and can be neglected.

To obtain the spin-orbit coupling Hamiltonian of chirons, the off-diagonal elements in Eq.~(6),
in the leading order in the interactions strength, it is sufficient to consider only the part of the
interaction Hamiltonian, $\hat H^{q=\pm2}_{ij}$, Eq.~(5),
which transfers momentum between the orbital
and rotational degrees of freedom of the molecules,
corresponding to the interaction propagator
$D(\bq)=D^{\mathrm{A}}(\bq)=D^{\mathrm{R}}(\bq)=(\pi/2)[(q^-)^2\ket{\uparrow}\bra{\downarrow}
+(q^+)^2\ket{\downarrow}\bra{\uparrow}]/q$.
As we have shown, the rest of the interaction Hamiltonian is responsible only for a slight
renormalisation of the kinetic energy $\xi_\bk$ of the molecules and
does not affect the spin-orbital coupling $\alpha(q)$.

The spin-orbital interactions couple the internal rotational states $\ket{\uparrow}$ and $\ket{\downarrow}$,
cf. Eq.~(6), and are characterised by the self-energy part
\begin{align}
	\Sigma^{\mathrm{A,R}}(\bk)&=\Sigma(\bk)
	\nonumber
	\\
	&=
	\frac{|\bd|^2}{6}\int_\bq D(\bq)
	\left[
	1\pm 2f_{00}(\bq+\bk)
	\right](\ket{\uparrow}\bra{\downarrow}
	+\ket{\downarrow}\bra{\uparrow}).
	\label{Sigma}
\end{align}
The first term in the square brackets in Eq.~(\ref{Sigma}) vanishes upon integration,
while the second term yields Eq.~(11).
When deriving the excitation spectrum and the spin-orbital interactions, Eq.~(\ref{Sigma}),
we assume that there is no phase coherence between the $\sigma=\ket{0,0}$ and $\sigma=\ket{1,\pm 1}$ states
in the background of the propagating chirons, and, therefore, $f_{0\uparrow}=f_{0\downarrow}=0$.

The effective Hamiltonian of the excitations is given by
\begin{equation}
	\hh(\bk)=\xi_\bk+\Sigma(\bk).
\end{equation}

\subsection*{Dynamics of two rotational excitations}

For two rotational excitations in a deep optical lattice, the ansatz
\begin{align}
\langle \mathbf{r}_{\mathbf{i}_1},\mathbf{r}_{\mathbf{i}_2}|\mathbf{K}\eta\rangle&=\frac{1}{L^2}\sum_{\mathbf{i}_1,\mathbf{i}_2}\sum_{\sigma_1\sigma_2}e^{-i\mathbf{K}\cdot (\mathbf{r}_{\mathbf{i}_1}+\mathbf{r}_{\mathbf{i}_2})}
\psi_{\mathbf{K};\sigma_1\sigma_2}^{\eta}\left(\mathbf{r}_{\mathbf{i}_1}-\mathbf{r}_{\mathbf{i}_2}\right)\hat{b}_{\mathbf{i}_1\sigma_1}^{\dagger}\hat{b}_{\mathbf{i}_2\sigma_2}^{\dagger}|\mathrm{vac}.\rangle\, ,
\end{align}
separates the two-body problem into a set of one-body problems for the relative coordinate functions $\psi^{\eta}_{\mathbf{K};\sigma\sigma'}\left(\mathbf{r}\right)$; one for each value of $\mathbf{K}$ in the Brillouin zone.  This reduces the computational scaling of direct diagonalisation of the two-excitation problem from $\mathcal{O}\left(L^{12}\right)$ to $\mathcal{O}\left(L^8\right)$.  The hard-core constraint breaks translational invariance in the relative degrees of freedom, and so direct numerical diagonalisation is required for the dynamics.

In order to verify that hard-core chiron-chiron interactions do not qualitatively affect the dynamics, we solved the two-excitation dynamics numerically in a variety of scenarios, using the Hamiltonian
\begin{align}
  \hat{H}&=-{J_0}\sum_{\mathbf{i}\ne \mathbf{j};\sigma}\frac{\hat{b}_{\mathbf{i}\sigma}^{\dagger}\hat{b}_{\mathbf{j}\sigma}}{ |\br_\mathbf{i}-\br_\mathbf{j}|^3}-{J_2}\sum_{\mathbf{i}\ne \mathbf{j}}\frac{e^{-2i\phi_{ij}}\hat{b}_{\mathbf{i}\uparrow}^{\dagger}\hat{b}_{\mathbf{j}\downarrow}+
e^{2i\phi_{ij}}\hat{b}_{\mathbf{i}\downarrow}^{\dagger}\hat{b}_{\mathbf{j}\uparrow}}{|\br_\mathbf{i}-\br_\mathbf{j}|^3}
+V\sum_{\bf i}\hat{n}_\mathbf{i}(\hat{n}_\mathbf{i}-1)\, .\label{eq:HCHami2}
\end{align}
Our results are collected in Figures~\ref{fig:Supp1}-\ref{fig:Supp2}.  In Figure~\ref{fig:Supp1} we show the dynamics in the scenario where the rotational wavepackets have a Gaussian spatial character and an $|\uparrow\rangle$ internal character.  In the top row, the initial wavepackets of the two rotational excitations overlap, i.e.~$\langle \mathbf{r},\mathbf{r} |\psi\rangle\ne 0$, and there are no interactions, $V=0$, corresponding to independent, non-interacting chirons.  The middle row removes any overlap between the two initial wavepackets, but still considers noninteracting dynamics.  The bottom row uses non-interacting wavepackets and the hard-core constraint, $V=\infty$, as is expected for experiments in a deep optical lattice.  The left and right sides of the panel show very short times and intermediate times.  The differences between the three instances, and especially between the latter two instances, are extraordinarily slight, justifying our neglect of chiron-chiron interactions in the main text.

In order to estimate the role of chiron-chiron interactions at higher densities, we also simulated the two-chiron dynamics in the case that the two rotational excitations are initially localised on two neighbouring lattice sites, see Figure~\ref{fig:Supp2}.  The top panel shows the non-interacting case ($V=0$), while the bottom row corresponds to hard-core interactions ($V=\infty$).  As in Figure~\ref{fig:Supp1}, differences in the two cases are slight.

\section*{Supplementary references}

\end{document}